\documentclass[%
 reprint,
 amsmath,amssymb,
 aps,twocolumn
]{revtex4}

\usepackage{graphicx}
\usepackage{dcolumn}
\usepackage{bm}

\usepackage{color}

\newcommand{\kenji}[1]{\textcolor{black}{ #1}}
\newcommand{\bukuro}[1]{\textcolor{black}{#1}}

\newcommand{\ichiki}[1]{\textcolor{black}{#1}}

\begin{document}


\title{Constraining the nature of ultra light dark matter particles with 21cm forest
}

\author{Hayato Shimabukuro}
 \affiliation{Department of Astronomy, Tsinghua Center for Astrophysics, Tsinghua University, Beijing 100084,China.}
 \email{bukuro@tsinghua.edu.cn}
\author{Kiyotomo Ichiki}
 \affiliation{%
Graduate School of Science, Division of Particle and Astrophysical Science, Nagoya University, Chikusa-ku, Nagoya, 464-8602, Japan
}%
 \affiliation{%
 Kobayashi-Maskawa Institute for the Origin of Particles and the Universe, Nagoya University, Chikusa-ku, Nagoya, 464-8602, Japan
}%
\author{Kenji Kadota}
\affiliation{
 Center for Theoretical Physics of the Universe, Institute for Basic Science (IBS), Daejeon, 34051, Korea
}%

\date{\today}

\begin{abstract}

The ultra-light scalar fields can arise ubiquitously, for instance, as a result of the spontaneous breaking of an approximate symmetry such as the axion and more generally the axion-like particles. In addition to the particle physics motivations, these particles can also play a major role in cosmology by contributing to dark matter abundance and affecting the structure formation at sub-Mpc scales. 
In this paper, we propose to use the 21cm forest observations to probe the nature of ultra-light dark matter.  The 21cm forest is the system of narrow absorption lines appearing in the spectra of high redshift background sources due to the intervening neutral hydrogen atoms, similar to the Lyman-$\alpha$ forest. Such features are expected to be caused by the dense neutral hydrogen atoms in a small starless collapsed object called minihalo.  The 21cm forest can probe much smaller scales than the Lyman-$\alpha$ forest, that is, $k\gtrsim 10 \mathrm{Mpc}^{-1}$. 
We explore the range of the ultra-light dark matter mass $m_{u}$ and $f_u$, the fraction of ultra-light dark matter with respect to the total matter, which can be probed by the 21cm forest. 
We find that 21cm forest can potentially put the dark matter mass lower bound $m_u \gtrsim 10^{-18}$ eV for $f_u=1$, which is 3 orders of magnitude bigger mass scale than those probed by the current Lyman-$\alpha$ forest observations.
While the effects of the ultra-light particles on the structure formation become smaller when the dominant component of dark matter is composed of the conventional cold dark matter, we find that the 21cm forest is still powerful enough to probe the sub-component ultra-light dark matter mass up to the order of $10^{-19}$ eV. The Fisher matrix analysis shows that $(m_u,f_u)\sim (10^{-20}\mathrm{eV}, 0.3)$ is the most optimal parameter set which the 21cm forest can probe with the minimal errors for a sub-component ultra-light dark matter scenario.
 
\end{abstract}

\maketitle

\section{Introduction}

Over the last decades, cosmological observations have provided us with a wealth of information on structure formation and evolution of the Universe.  In particular, the fluctuations of cosmic microwave background (CMB) observed by WMAP and Planck satellites and the matter density fluctuations on large scale structure (LSS) revealed that most current observations are consistent with the $\Lambda$CDM cosmological scenario based on the cold dark matter (CDM), cosmological constant and inflation model\citep[e.g.][]{2011ApJS..192...18K,2015PhRvD..92l3516A,2016A&A...594A..13P,2018PhRvD..98d3526A}. The $\Lambda$CDM cosmological model has been a concrete framework well describing the Universe at larger scales. However, focusing on scales smaller than $\sim$1 Mpc, the numerical simulations based on the $\Lambda$CDM model face the apparent disagreement with the observations such as ``missing satellite problem''\citep[e.g.][]{1999ApJ...524L..19M},``core-cusp problem''\citep[e.g.][]{2010AdAst2010E...5D}, and ``Too big to fail problem'' \citep[e.g.][]{2012MNRAS.422.1203B,2017ARA&A..55..343B}. As a prescription to understand these discrepancies, the properties of the dark matter have been explored beyond the simple CDM model which can possibly suppress the small scale structures such as the ultra-light scalar dark matter.

The light scalar fields can commonly arise, for instance, as a result of the spontaneous breaking of an approximate symmetry such as the axion and more generally the axion-like particle, and many experiments have been searching for such a light scalar field which can also contribute to the dark matter abundance \cite{Peccei:1977hh,Weinberg:1977ma,Wilczek:1977pj,sik1983,as2009,ana2017,raf1987,kel2017,hua2018,hook2018,har1992,fed2019,kad2019}. While the mass of such a light particle can span a wide range with a heavy model dependence, the ultra-light mass ($\lesssim 10^{-20}$eV) is well motivated from the particle theory (sometimes referred to as the string axiverse) and also of great interest from the cosmology because it can lead to the substructure suppression within the Jeans/de Broglie scale (often refereed to as the fuzzy dark matter) \cite{sv2006,Arvanitaki:2009fg, hu2000,ame2005,Hui:2016ltb,marsh15,Kadota:2013iya,sch2018}. 
One of the promising methods to probe the nature of ultra-light dark matter is 21cm observations. A neutral hydrogen atom emits or absorbs the radio wave, whose wavelength is 21cm in rest frame, due to its hyperfine structure\citep[e.g.][]{1990MNRAS.247..510S,1997ApJ...475..429M}. Since the 21cm signal is redshifted, we can tomographically probe the Universe by following the redshift evolution of the 21cm signals.  We typically focus on the 21cm emission line from hydrogen atom in the Intergalactic medium(IGM) to study the thermal and ionized states of the IGM during the epoch of reionization (EoR)\citep[e.g.][]{2006PhR...433..181F,2012RPPh...75h6901P}. At a high redshift during the EoR and beyond, the smallest bound objects called ``minihalos'' can form which have the virial temperature below the threshold where atomic cooling becomes effective($T_{\mathrm{vir}} \lesssim 10^{4}\mathrm{K}$ or $M \lesssim 10^{8}M_{\odot}$). They thus cannot cool effectively and cannot collapse to form proto-galaxies.  The neutral hydrogen atoms in such a minihalo generate 21cm absorption lines in the continuum emission spectrum from the high redshift luminous radio background sources such as radio quasars and gamma-ray bursts (GRBs) \citep[e.g.][]{2015aska.confE...6C}.  We call the system of these 21cm absorption lines `` 21cm forest'' in analogy with the Lyman alpha forest\citep[e.g.][]{2002ApJ...577...22C,2002ApJ...579....1F,2006MNRAS.370.1867F}. The mass scale of the minihalos corresponds to $k \gtrsim 10[\mathrm{Mpc}^{-1}]$, which is much smaller than the scales Lyman-$\alpha$ forest can probe \cite{Chabanier:2018rga,2015aska.confE...6C}. A wide range of the mass scale is available for the ultra-light dark matter \cite{Arvanitaki:2009fg} and consequently the scales where the matter fluctuation suppression show up can go well below the currently accessible scales such as those by Lyman-$\alpha$ observation, and the 21cm forest can potentially offer a unique mean to study the properties of ultra-light dark matter.We expect that 21cm forest can be accessible with future observations such as Square Kilometre Array (SKA) as long as sufficiently bright radio sources exist at a relevant redshift ($z\gtrsim 6$)\cite{2015aska.confE...6C}.

The properties of the ultra-light dark matter can be characterized by its mass $m_u$ and the abundance $f_{u}=\Omega_{u}/\Omega_{m}$ which represents the fraction of ultralight dark matter with respect to the total matter abundance. The range of the parameter values which can be explored depends on the specifications of the experiments under consideration. For instance, the CMB which can probe the linear scale can explore the mass range $m_{u} \lesssim 10^{-26}\mathrm{eV}$ for $f_{u}$ $\gtrsim 0.01$\cite{2017PhRvD..95l3511H}. The exploration for the larger mass range requires the sensitivity to the smaller scale \cite{Kadota:2013iya, 2017PhRvL.119c1302I}, and the Lyman-$\alpha$ for instance can exclude $m_u\lesssim 10^{-22}$eV for $f_u=1$. We will demonstrate that the 21cm forest can be a unique probe on the ultra light particles which can be sensitive to the mass up to $10^{-18}$ eV which is not amenable to any other experiments.

For concreteness, to model the ultra-light dark matter, we consider a ultra-light scalar field $\phi$ in a quadratic potential for which $\phi$ behaves as a dark energy component due to the Hubble friction until $H(t)$ equals $m_u$ and behaves as a dark matter component afterwards. The ultra-light particles can contribute to the current dark matter abundance for $m_u> H_0 \sim 10^{-33}$ eV due to the coherent oscillations and we implement such an ultra-light scalar field in the publicly available Boltzmann code CAMB \cite{Lewis:1999bs}.

\section{Matter power spectrum}
\label{sec2}

\begin{figure*}[htbp]
\includegraphics[width=0.8\hsize]{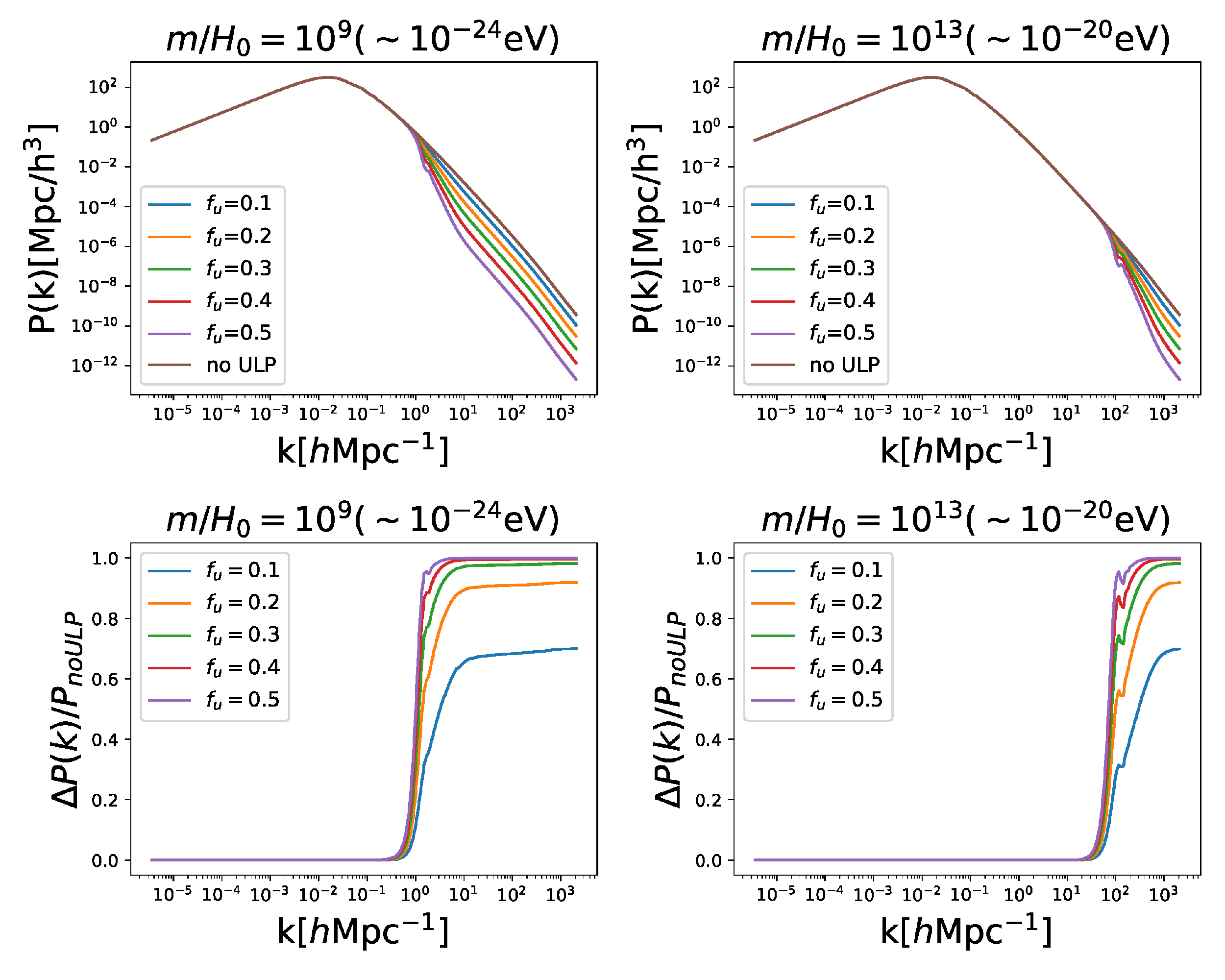}
\caption{({\it top})  The effect of ULPs on matter power spectrum for different ULP abundances $f_u=\Omega_u/\Omega_{\mathrm{DM}}$ (for the ULP masses $m/H_{0}=10^{9}, 10^{13}$).  ({\it bottom}). The relative change in the matter power spectrum, defined by eq.(\ref{eq:rel_err}).}
\label{fig:pk_ratio1}
\end{figure*}

\begin{figure*}[htbp]
\includegraphics[width=0.8\hsize]{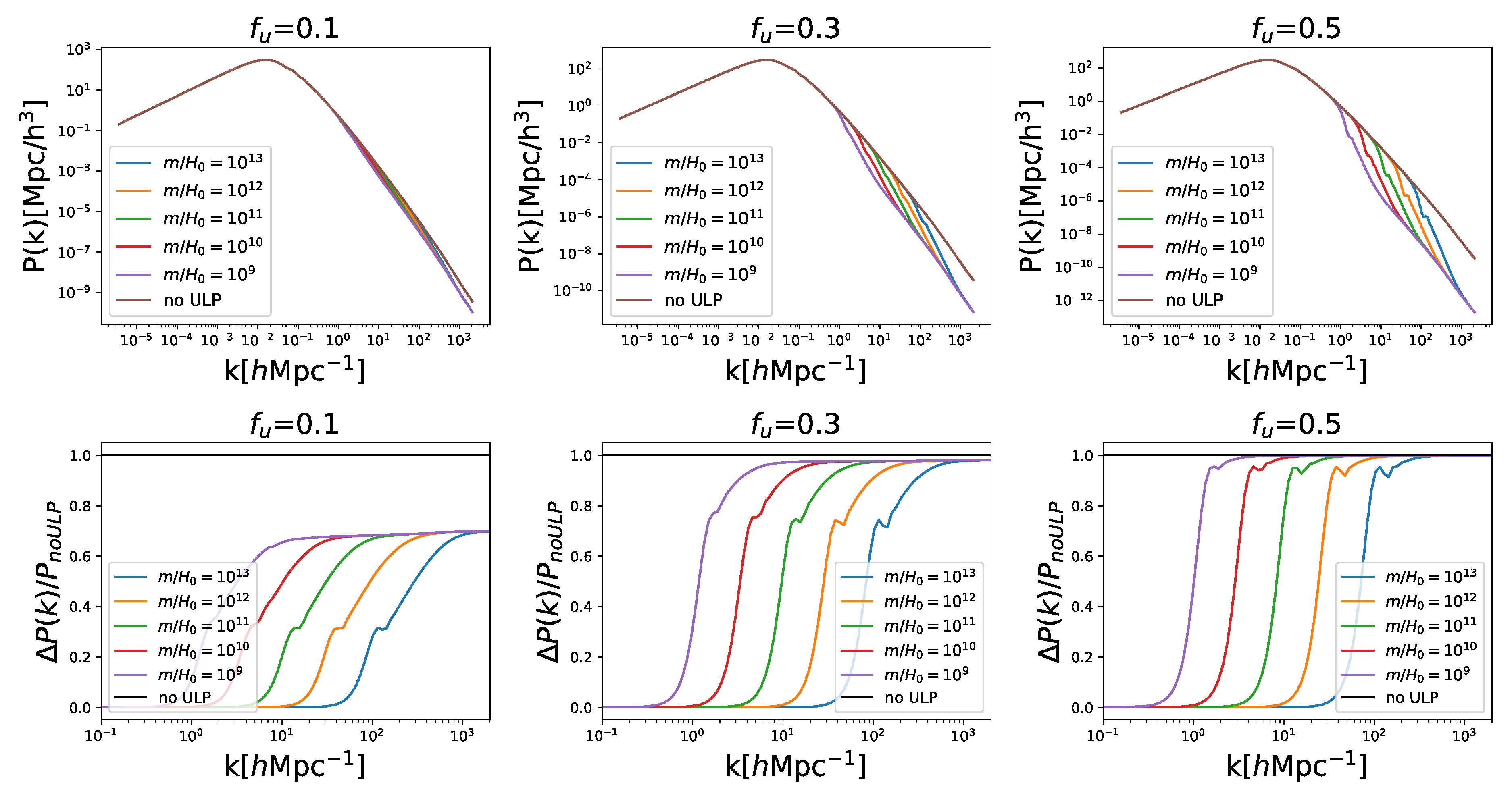}
\caption{({\it top}) The effect of ULPs on matter power spectrum for the different ULP masses for a given $f_u$. ({\it bottom}) The corresponding relative change of the matter power spectrum.}
\label{fig:pk_ratio2}
\end{figure*}

We first see the effect of ultra-light particles (ULPs) on the matter power spectrum. 
The power spectrum dependence on our parameters $m_u,\Omega_u$ is illustrated in Figs. \ref{fig:pk_ratio1} and \ref{fig:pk_ratio2}. The top figures in Fig.\ref{fig:pk_ratio1} vary the fraction of ULPs while fixing the mass of ULPs, and the bottom figures show the corresponding relative change of the matter power spectrum between CDM model and ULP model defined as

\begin{equation}
    \frac{\Delta P(k)}{P(k)}=\frac{P_{\mathrm{no ULP}}(k)-P_{\mathrm{ULP}}(k)}{P_{\mathrm{no ULP}}}
    \label{eq:rel_err}
\end{equation}
Meanwhile, we also show the matter power spectrum varying the ULP mass $m_u$ while fixing the fraction $f_u$ in Fig.\ref{fig:pk_ratio2}. 

As can be seen from these figures, the mass of ULP controls the scale of suppression and the fraction of ULP affects how much the amplitude of the matter power spectrum is suppressed.  The less massive ULPs have bigger Jeans/de Broglie length scales than the massive ones and thus can suppress the matter fluctuations at larger scales, and the amplitude suppression becomes bigger for a bigger ULP abundance (the total amount of dark matter (ULP plus the conventional CDM) is fixed). For the linear scale, analogously to the matter power spectrum suppression due to the neutrinos, $\Delta P/P \propto \Omega_u/\Omega_m$ \cite{bond1980}. From Fig.\ref{fig:pk_ratio2}, we also find that matter power spectrum is independent of the ULP mass at a high $k$. It is because the fluctuations entering the horizon during the radiation dominant epoch cannot grow. Furthermore, the fluctuations inside the Jeans scale cannot grow regardless of ULP mass even during the matter dominant epoch. This leads to ULP's mass independent matter power spectrum at a sufficiently high $k$. 

More quantitatively, the dependence of the scale where the suppression shows up on the ultra-light particle mass $m_u$ can be inferred from the Jeans scale inside which the pressure prevents the matter fluctuation growth. The Jeans length scales as $\lambda_{\rm J}\sim c_s /\sqrt{G \rho}$ (representing the length scale which a sound wave with the speed $c_s$ travels within the dynamical time scale for the free fall collapse $\tau_{\rm grav}\sim (G \rho)^{-1/2}$ ($\rho$ is the total energy density), where a characteristic feature for the ultra-light particle is the scale dependent sound speed $c_s \approx k/2m_u$ for $a\gg k/2m_u$ and $c_s \approx 1$ for $a\ll k/2m_u$ \cite{hu2000,hw2009}. When the ultra-light scalar field starts oscillation during the matter domination at $z_{{\rm osc}}\sim (m_u^2/H_0^2 \Omega_m)^{1/3}$ corresponding to $H(z_{{\rm osc}})\sim m_u$, the matter power spectrum would be suppressed for $k \gtrsim k_{\rm J}$ where $k_{\rm J}$ is the Jeans scale scaling as $k_{\rm J} \sim \left( H_0^2 \Omega_m   \right )^{1/3} m_u^{1/3}$. Analogously, when the oscillation starts during the radiation domination (which is the case for the parameter range of our interest in this paper), the suppression occurs for the scales smaller than the Jeans scale at the matter-radiation equality $k_{\rm J} \sim \left( m_u^2 H_0^2 \Omega_m a_{eq}\right )^{1/4} \sim 3 \times 10^{-5} (m_u/H_0)^{1/2} [h$Mpc$^{-1}]$ which gives a reasonable estimation to our numerical results as shown in Figs. \ref{fig:pk_ratio1} and \ref{fig:pk_ratio2} \cite{hu2000,ame2005,Arvanitaki:2009fg}.

\section {Halo and gas profile}\label{sec:halo_gas}

Our basic formulation follows that given by \cite{2002ApJ...579....1F} and \cite{2001PhR...349..125B} with due modifications for our purposes. We start with the description of the gas density profile in dark matter halos.
We assume that the dark matter potential is described by the
Navarro, Frenk $\&$ White (NFW) profile \cite{1997ApJ...490..493N,2000ApJ...540...39A} characterized by the concentration parameter ${\it y}=r_{{\rm vir}}/r_{{\rm s}}$, where $r_{s}$ is the scaling radius and the virial radius
$r_{{\rm vir}}$ is given by \cite{2001PhR...349..125B}
\begin{multline}
  r_{{\rm vir}}=0.784\bigg(\frac{M}{10^{8}h^{-1}M_{\odot}}\bigg)^{1/3}\bigg[\frac{\Omega_{m}}{\Omega_{m}^{z}}
  \frac{\Delta_{c}} {18\pi^{2}}\bigg]^{-1/3}\\
  \times \bigg(\frac{1+{\it z}}{10}\bigg)^{-1}h^{-1}[{\rm kpc}]
\label{eq:virial_radius}
\end{multline}
where $\Delta_{c}=18\pi ^{2}+82d-39d^{2}$ is the overdensity of halos collapsing at redshift ${\it z}$,
with $d=\Omega_{m}^{z}-1$ and
$\Omega_{m}^{z}=\Omega_{m}(1+z)^{3}/(\Omega_{m}(1+z)^{3}+\Omega_{\Lambda})$. 
Here we follow the N-body simulation results of \citet{2005MNRAS.363..379G} for halos at high-redshift and assume that $y$ is inversely proportional to $(1+z)$.
Within the dark matter halo, the gas is assumed to be isothermal and in hydrostatic equilibrium, for which its profile can be derived analytically \cite{1998ApJ...497..555M,2011MNRAS.410.2025X}.
The gas density profile is given by
\begin{equation}
  \ln \rho_{{\rm g}}(r)=\ln \rho_{{\rm g0}}-\frac{\mu m_{{\rm p}}}{2k_{{\rm B}}T_{{\rm vir}}}[v_{{\rm esc}}^{2}(0)-v_{{\rm esc}}^{2}(r)],
\label{eq:gas_profile1}
\end{equation}\\
where
\begin{multline}
  T_{{\rm vir}} =1.98\times 10^{4}\bigg(\frac{\mu}{0.6}\bigg)\bigg(\frac{M}{10^{8}h^{-1}M_{\odot}}\bigg)^{2/3}\\
  \times \bigg[\frac{\Omega_{m}}{\Omega_{m}^{z}}\frac{\Delta_{c}}{18\pi^{2}}\bigg]^{1/3}\bigg(\frac{1+z}{10}\bigg)[{\rm K}]
\end{multline}
is the virial temperature,
$\rho_{{\rm g0}}$ is the central gas density, $m_{{\rm p}}$ is the
proton mass and $\mu=1.22$ is the mean molecular weight of the gas.
The escape velocity $v_{{\rm esc}}(r)$ is described by
\begin{equation}
  v_{{\rm esc}}^{2}(r) = 2\int_{r}^{\infty}\frac{GM(r^{'})}{r^{'2}}dr^{'}=2V_{c}^{2}\frac{F(yx)+yx/(1+yx)}{xF(y)}~,
\label{eq:esc_velocity}
\end{equation}
where $x\equiv r/r_{{\rm vir}}$ and $F(y)=\ln(1+y)-y/(1+y)$, and
$V_{c}$ is the circular velocity given by
\begin{multline}
  V_{c}^{2} =  \frac{GM}{r_{{\rm vir}}}=23.4\bigg(\frac{M}{10^{8}h^{-1}M_{\odot}}\bigg)^{1/3}
  \bigg[\frac{\Omega_{m}}{\Omega_{m}^{z}}\frac{\Delta_{c}}{18\pi^{2}}\bigg]^{1/6}\\
  \times \bigg(\frac{1+z}{10}\bigg)^{1/2}[{\rm km/s}].
  \label{eq:cir_velocity}
\end{multline}

The escape velocity reaches its maximum of $v_{{\rm esc}}^{2}(0)=2V_{c}^{2}y/F(y)$ at the center of the halo.
The central density $\rho_{g0}$ is normalized by the cosmic value of $\Omega_{b}/\Omega_{m}$ and given by

\begin{equation}
  \rho_{g0}(z) =
  \frac{(\Delta_c/3)y^{3}e^{A}}{\int_{0}^{y}(1+t)^{A/t}t^{2}dt} \left(\frac{\Omega_b}{\Omega_m}\right)\bar{\rho}_{m}(z)~,
\end{equation}
where $A=3y/F(y)$ and $\bar{\rho}_{m}(z)$ is the mean total matter
density at redshift $z$.

\section{21cm forest}

In order to evaluate the 21cm forest quantitatively, we need to compute the 21cm optical depth to 21cm absorption.  The 21cm optical depth is characterized by the spin temperature $T_{\mathrm{S}}$ and HI column density in a halo.  We briefly outline how we treat these quantities in the following subsections (see \cite{2014PhRvD..90h3003S} for the further details.)

\subsection{Spin temperature}
The spin temperature describes the excitation state of the hyperfine transition in the HI atom.  The spin temperature is generally determined by the interaction between HI atom and CMB photons, collisions of HI atom with other particles by the following equation:
\begin{equation}
  T_{{\rm S}}^{-1}=\frac{T_{\gamma}^{-1}+x_{c}T_{{\rm K}}^{-1}+x_{\alpha}T_{{\rm C}}^{-1}}{1+x_{c}+x_{\alpha}}.
\label{eq:spin1}
\end{equation}
Here, $T_{\gamma}=2.73(1+z)$ is CMB temperature at redshift $z$, $T_{\mathrm{K}}$ is the gas kinetic temperature and $T_{\mathrm{C}}$ is color temperature of UV radiation field. $x_{\alpha}$ and $x_{{c}}$ are the coupling coefficients for collision with UV photons and other particles, respectively. In this work, we ignore any UV radiation field and radiative feedback in order to better understand how the 21cm forest is affected by the modification of cosmological effect. Thus, we set $x_{\alpha}=0$. We also set $T_{\mathrm{K}}=T_{\mathrm{vir}}$, which should be a good approximation for the minihalos in which the gas cooling is inefficient.  For computation of $x_{c}$, we take into account the coupling coefficient for H-H interaction \cite{2005ApJ...622.1356Z,2006MNRAS.370.1867F,2014PhRvD..90h3003S}. The spin temperature approaches the virial temperature (hence a larger halo has a larger spin temperature) in the inner regions of minihalos and approaches the CMB temperature in the outer regions where the collisional coupling becomes ineffective due to a small gas density \cite{2014PhRvD..90h3003S}. 

\subsection{Optical depth}
The optical depth to 21cm absorption by the neutral hydrogen gas in a minihalo of mass $M$ (at a frequency $\nu$ and at an impact parameter $\alpha$) is given by 
\begin{multline}
  \tau(\nu,M,\alpha)=\frac{3h_{{\rm p}}c^{3}A_{10}}{32\pi k_{{\rm B}}\nu_{21}^{2}}
  \int_{-R_{{\rm max}}(\alpha)}^{R_{{\rm max}}(\alpha)}dR\frac{n_{{\rm H{\sc I}}}(r)}{T_{{\rm S}}(r)\sqrt{\pi}b}\\
  \times \exp\bigg(-\frac{v^{2}(\nu)}{b^{2}}\bigg),
\label{eq:optical1}
\end{multline}
where the velocity dispersion $b=\sqrt{2k_{B}T_{{\rm vir}}/m_p}$ and $R_{\mathrm{max}}$ is the maximum radius of halo at $\alpha$. $n_{\mathrm{HI}}$ is the number density of neural hydrogen gas in a halo.  \bukuro{Note that we only consider thermal broadening and ignore the infall velocity  around minihaloes projected along the line of sight.}As seen in section \ref{sec:halo_gas}, we assume the neutral hydrogen gas is isothermal and in hydrostatic equilibrium within dark matter halo. A smaller impact parameter results in a larger optical depth due to a larger column density despite of a larger spin temperature, and a smaller mass leads to a smaller spin temperature hence a larger optical depth at a fixed impact parameter (e.g. Fig.2 in \citet{2014PhRvD..90h3003S}).

\subsection{Abundance of 21cm absorbers}
We introduce the following function to represent the abundance of the 21cm absorption lines

\begin{equation}
  \frac{dN(>\tau)}{dz}=\frac{dr}{dz}\int_{M_{{\rm min}}}^{M_{{\rm max}}}dM\frac{dN}{dM}\pi r^{2}_{\tau}(M,\tau), 
  \label{eq:abundance1}
\end{equation}
where $dr/dz$ is the comoving line element, $r_{\tau}(M,\tau)$ is the maximum impact parameter in comoving units that gives the optical depths greater than $\tau$,and $dN/dM$ is the halo mass function
representing the comoving number density of collapsed dark matter halos with mass between $M$ and $M+dM$, here given by the Press-Schechter formalism\cite{1974ApJ...187..425P}.  \bukuro{We note that Sheth-Tormen mass function is more precise at least for a low redshift \cite{1999MNRAS.308..119S}. However, the situation is currently less clear for high redshifts of our interest and we actually checked that the effect of difference between Press-Schecther formalism and Sheth-Tormen formalism is small. Thus we use Press-Schechter formalism in this work.}  The maximum mass $M_{{\rm max}}$ for minihalos is determined by the mass with $T_{{\rm vir}}=10^{4}$ K, below which the gas cooling via atomic transitions and the consequent star formation is expected to be inefficient. \bukuro{Instead of atomic cooling, molecular hydrogen cooling may allow star formation in smaller halos.  However, it is expected that molecular hydrogen  is photo-dissociated by Lyman-Werner radiation background emitted by stars\citep[e.g.][]{2001ApJ...548..509M,2007ApJ...671.1559W}.} The minimum mass $M_{{\rm min}}$ is assumed to be the Jeans mass determined by the IGM temperature $T_{\mathrm{IGM}} $\cite{2003tsra.symp...31M}

\begin{equation}
\begin{split}
  M_{{\rm J}} &=\frac{4\pi \bar{\rho}}{3}\left(\frac{5\pi k_{{\rm B}}T_{\rm IGM}}{3G\bar{\rho}m_{{\rm p}}\mu}\right)^{3/2}\\
  &\simeq 3.58 \times 10^{5}h^{-1}M_{\odot}\left(\frac{T_{\rm IGM}/{\rm K}}{1+z}\right)^{3/2}~,
\label{eq:jeans}
\end{split}
\end{equation}
where $\bar{\rho}$ is the total mass density including dark matter. We choose $T_{\mathrm{IGM}}=10\times T_{\mathrm{ad}}$,10 times of the average temperature of the IGM assuming the adiabatic cooling by cosmic expansion. This assumption is valid in our case because we do not account the astrophysical radiative feedback.\bukuro{Note that the time-averaged filtering mass is a better mass scale to be used than the Jeans mass because it takes the time evolution of gas to respond to earlier heating into account\citep[e.g.][]{1998MNRAS.296...44G,2011MNRAS.418..906T}. This makes it possible to follow prior thermal history. Typically, the filtering mass is lower than the Jeans mass. However, we adopt the Jeans mass for simplicity.}

\section{Results}
\subsection{The abundance of 21cm absorption lines}

\begin{figure*}[htbp]

\includegraphics[width=1.0\hsize]{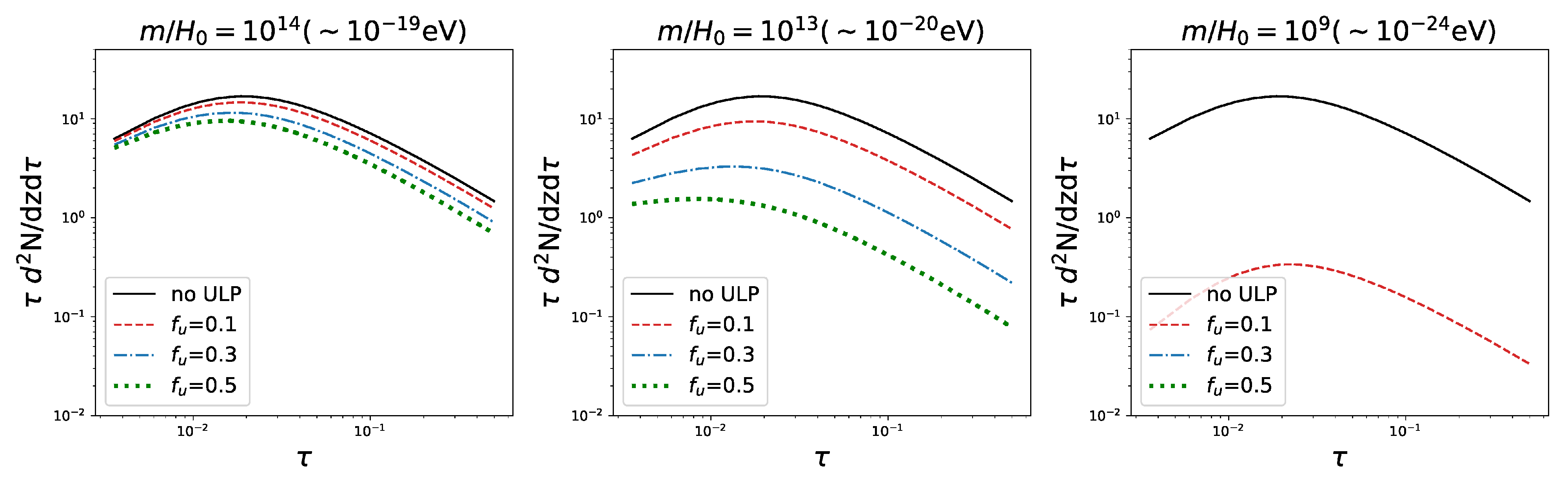}
\includegraphics[width=1.0\hsize]{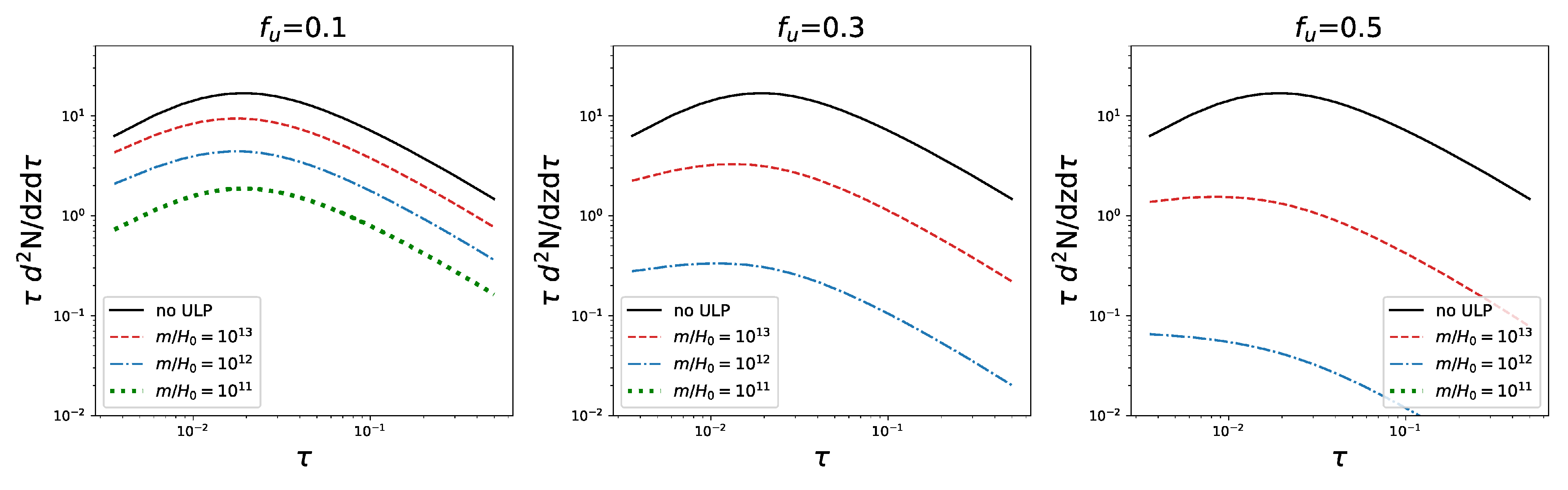}
\caption{({\it top}) The abundance of 21cm absorption lines as a function of optical depth $\tau$ for different ULP abundances ($f_{u}=0.1,0.3,0.5$) and masses $m/H_{0}=10^{14},10^{13}$ and $10^{9}$. As a reference, we also show the 21cm absorption lines in the case of the conventional cold dark matter without the ULP ({\it solid}). At $m/H_{0}=10^{9}$, the number of 21cm absorption lines for $f_{u}=0.3,0.5$ is too small to be shown. ({\it bottom}) Same as the top figure, with different masses for a given ULP abundance. For $f_{u}=0.3$ and $0.5$, the number of 21cm absorption lines for $m/H_{0}=10^{11}$ is too small to be shown.}
\label{fig:21cm_abs}
\end{figure*}

In Fig.\ref{fig:21cm_abs}, we show the abundance of 21cm absorption lines at $z=10$ as a function of the optical depth per redshift interval along line of sight and per optical depth. At the top of Fig\ref{fig:21cm_abs}, we vary the abundance of ULPs for a given ULP mass, while the bottom figures vary the ULP mass for a given ULP abundance. We can see that a larger ULP abundance and a smaller ULP mass suppress more the abundance of the 21cm absorption lines, as expected by the effects of ULPs on matter power spectra demonstrated by Figs.\ref{fig:pk_ratio1} and \ref{fig:pk_ratio2}. We are interested in the minihalo mass range around $10^5 \sim 10^8 M_{\odot}$ (corresponding to $30\lesssim k \lesssim 10^{3}$), where the lower bound comes from the baryon Jeans mass and the upper bound from the insufficient atomic cooling, and those scales are indeed expected to be suppressed due to ULP whose Jeans scale approximately scales as $k_{\rm J}  \sim 3 \times 10^{-5} (m_u/H_0)^{1/2} [h$Mpc$^{-1}]$ for the parameter range of out interest as discussed in Section \ref{sec2}.

For a comparison, in the case of the conventional CDM model without ULPs, the expected number of absorption lines is $\mathcal{O}(10)$ around the peak optical depth $\tau \sim 0.1$. Depending on the fraction and mass of ULPs, we can see the number of 21cm absorption lines can be suppressed by more than an order of magnitude. Such a big change in the abundance can make it feasible for the 21cm forest observations to differentiate the different models. We, on the other hand, point out that Fig.\ref{fig:21cm_abs} illustrates that it is challenging to detect 21cm absorption lines for too large values of $f_u$ and too small values of $m_u$ because of too much suppression. For instance, in the case of $m/H_{0}\lesssim 10^{11}$, the number of 21cm absorption lines is less than $\mathcal{O}(1)$ at $f_{u} \gtrsim 0.1$. Such a small value of $m_u$ can be probed by other observations probing the larger scales such as the CMB and Lyman-$\alpha$, and the 21cm forest could be complementary to the other experimental constraints to study the previously unexplored parameter space of the ULPs.%

\begin{figure*}[htbp]

\includegraphics[width=1.0\hsize]{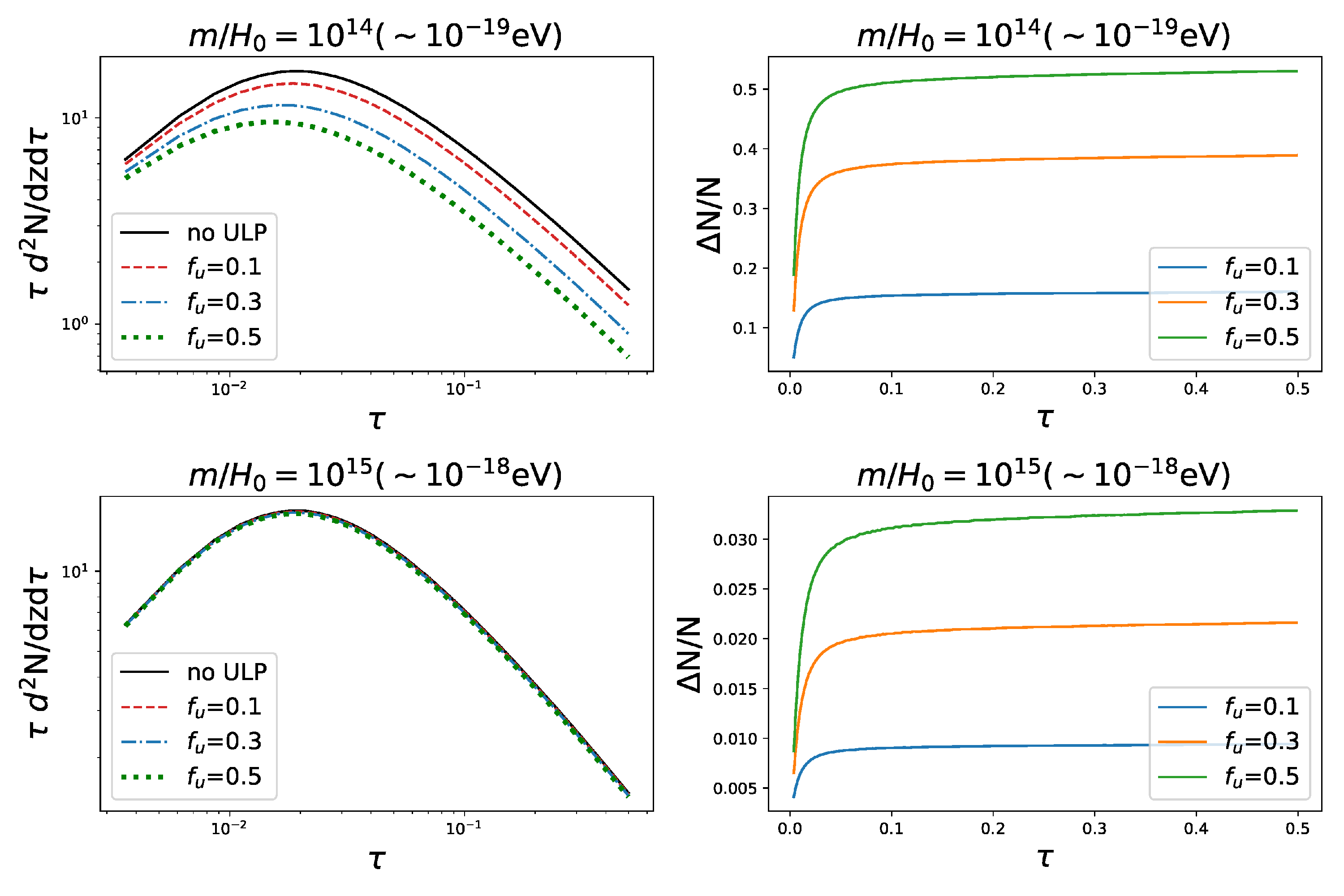}
\caption{({\it top}) The number of the 21cm absorption lines and the fractional difference from the conventional CDM model without ULPs.}
\label{fig:21cm_abs2}
\end{figure*}

More details also can be seen in Fig.\ref{fig:21cm_abs2} which also shows the fractional change with respect to the CDM model without ULP for different ULP abundances. In the case of $m/H_0=10^{14}$, we can see the reduction of more than 10\% in the number of 21cm absorption compared with no ULP model even with $f_u=0.1$. On the other hand, from bottom right panel of Fig.\ref{fig:21cm_abs2}, we can see less than only 0.3\% reduction in the number of the 21cm absorption lines for $m/H_0=10^{15}$ if $f_u=0.1$.  

\begin{figure}[htbp]

\includegraphics[width=1.0\hsize]{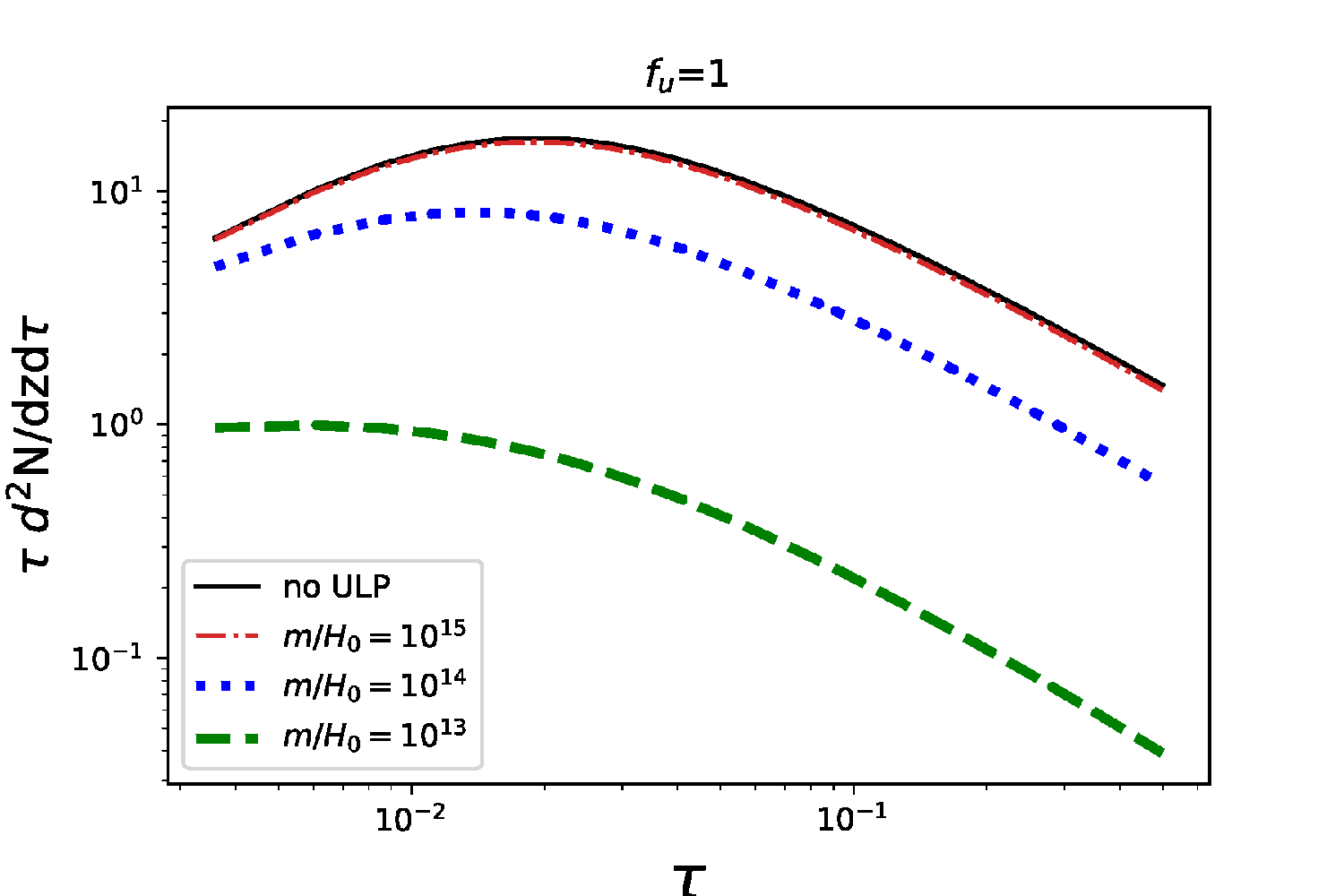}
\caption{The number of 21cm absorption lines in the case of $f_u$=1 varying $m/H_{0}=10^{13},10^{14},10^{15}$.}
\label{fig:21cm_abs3}
\end{figure}

In Fig.\ref{fig:21cm_abs3}, we illustrate the maximum ULP mass to be explored by the 21cm forest for $f_u=1$. We can infer that 21cm forest can explore $f_{u}=1$ case if ULP's mass is less than $m/H_0=10^{15}$ which is $\sim$ 3 order of magnitude higher mass scale than that currently probed by Lyman-$\alpha$ forest. 

We in the following section perform the Fisher matrix likelihood analysis to give a more quantitative estimation for the ULP parameter ranges which can be probed by the 21cm forest.

\subsection{Fisher analysis}

We assume that the number of absorption lines obey the Poisson
statistics \cite{1979ApJ...228..939C}. The Poisson distribution function is given by
\begin{equation}
 {\cal L}(n,\bar{n}) = \bar{n}^n e^{-\bar{n}} / n! ~,
\end{equation}
where $n$ is the number of absorption lines and $\bar{n}$ is its
expectation value for the fiducial model. 
The log-likelihood is then given by
\begin{equation}
 \ln{\cal L} = n\ln(\bar{n}) - \ln(n!)
\end{equation}
where we have omitted an irrelevant constant. 
The fisher matrix is defined as
\begin{equation}
 F_{ij}= -\frac{\partial^2 \ln{\cal L}}{\partial\theta_i
  \partial\theta_j}|_{\vec{\theta}=\vec{\theta}_{\rm fid}},
\end{equation}
where $\vec{\theta}$ is the parameter vector, and $\vec{\theta}_{\rm fid}$ denotes the fiducial parameters. 

We consider the number of absorption lines integrating over
optical depth $\tau$ for our statistics and it is given by
\begin{equation}
 n^{(\tau_i,z_j)} = \int_{\tau_i}^{\tau_i+\Delta \tau}
  \int_{z_j}^{z_j+\Delta z} \frac{d^2N}{d\tau dz} d\tau dz
  \label{eq:tau_integration}
\end{equation}
where the distribution $d^2N/d\tau dz$ is obtained by taking a derivative of Eq. \ref{eq:abundance1} with respect to $\tau$ and is shown in Fig.\ref{fig:21cm_abs}.

Here $n^{(\tau_i,z_j)}$ is the number in the optical depth in the bin at $(\tau_i,z_j)$ with the widths of $\Delta \tau$ and $\Delta z$ for the optical depth and redshift, respectively. We set 
the redshift interval $\Delta z=1$ for our estimate. The minimum optical
depth $\tau_{\rm min}$ should be determined by the sensitivity of
experiments. In the following analysis we divide the optical depth into two bins \ichiki{$[0.003,0.02]$ and $[0.02:1]$} to break the degeneracy between $f_u$ and $m_u$.
The total likelihood of observing
$n^{(\tau_i,z_j)}$ absorption lines in each $(\tau_i,z_j)$ bin can then be
written as 
\begin{equation}
 {\cal L} = \Pi_{\tau_i} \Pi_{z_j}
 {\cal L}(n^{(\tau_i,z_j)},\bar{n}^{(\tau_i,z_j)}) ~,
\end{equation}
and the log likelihood becomes
\begin{equation}
 \ln{\cal L} = \sum_{\tau_i}\sum_{z_j}
  n^{(\tau_i,z_j)}\ln(\bar{n}^{(\tau_i,z_j)}) - \ln(n^{(\tau_i,z_j)}!)~.
\end{equation}

\subsection{Fisher analysis result}
\begin{figure*}[h]
 \includegraphics[width=0.3\textwidth]{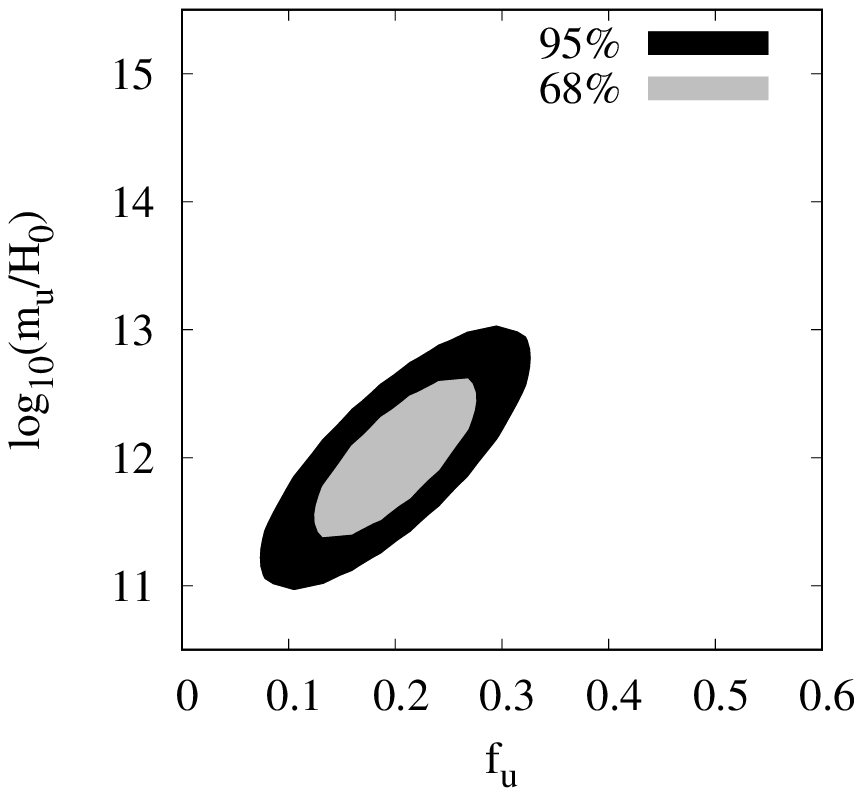}
 \includegraphics[width=0.3\textwidth]{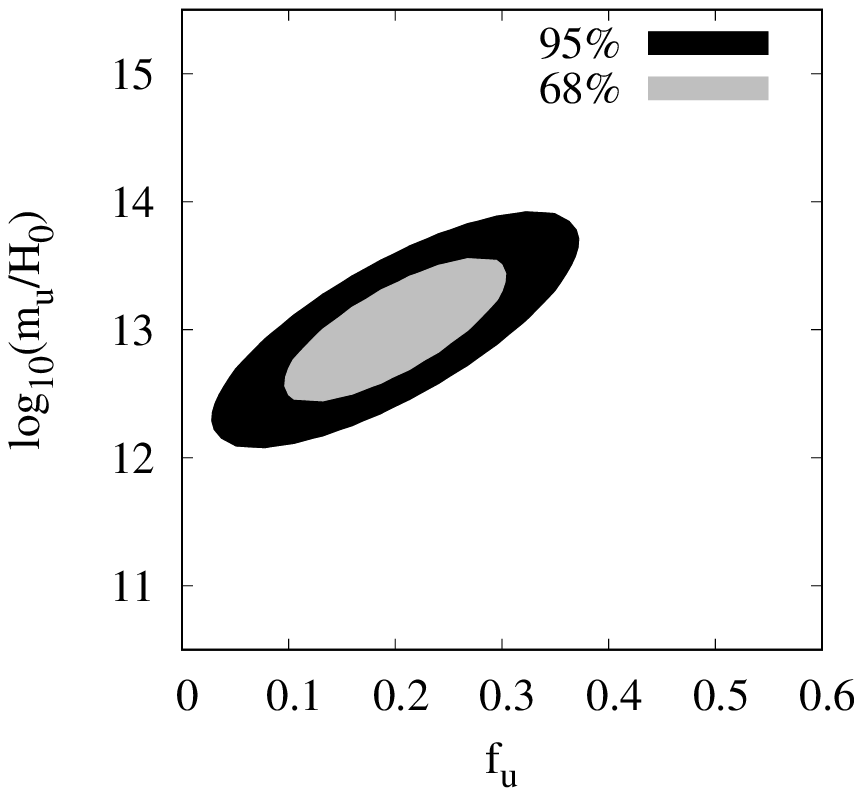}
 \includegraphics[width=0.3\textwidth]{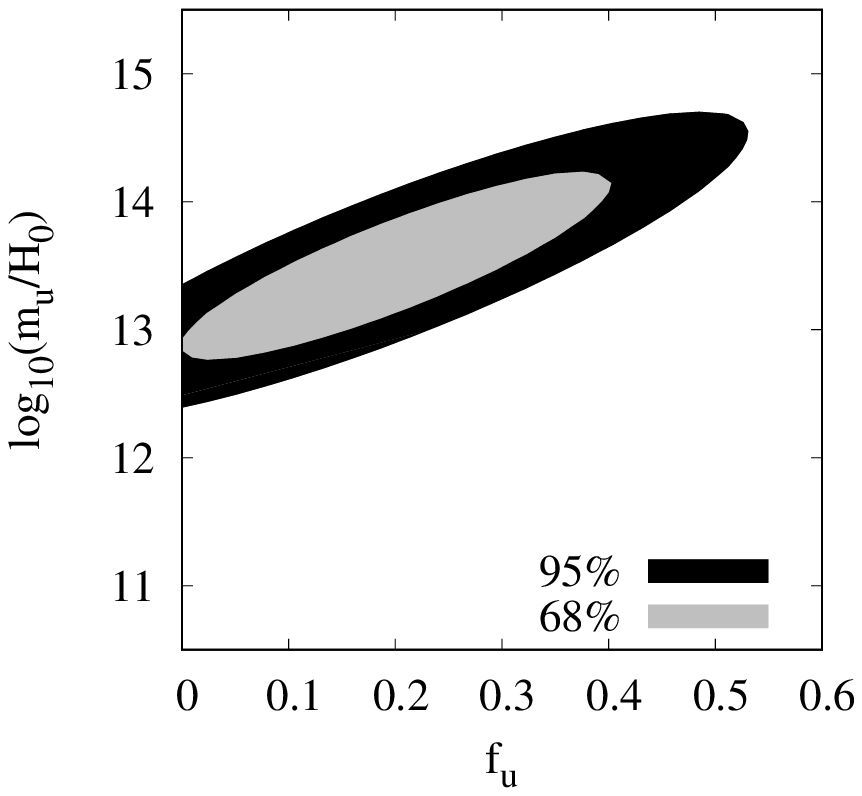}
 \caption{Expected constraints on the $f_u$ and $\log_{10}(m_u/H_0)$
 plane for $\log_{10}(m_u/H_0)=12.0$, $13.0$ and $13.5$ from left to
 right, respectively. The fiducial value of the ULP dark matter fraction
 is fixed to $f_u = 0.2$. For $\log_{10}(m_u/H_0)=14.0$ we can not
 obtain any sensible constraints.}  \label{fig:ma}
\end{figure*}

In Fig.~\ref{fig:ma}, we show the two dimensional constraints on
$\log_{10}(m_u/H_0)$ and $f_u$ for the fixed fiducial value $f_u=0.2$. We find, even with the reduced effects due to a small $f_u<1$, 
that the sensible constraints can be obtained even for the case with
$\log_{10}(m_u/H_0)=13.5$ owing to the sensitivity of the 21cm forest
observation on small scales. The error bars become larger for a larger $m_u$ because ${\delta N}/{N}$ (the fractional reduction in an absorption line abundance) becomes smaller.
For the case with $\log_{10}(m_u/H_0)=15.0$, we find that the
constraint ellipse does not close within the range of $0\le f_u \le1$
and we can not put any constrains on ULP parameters in this case because of too small a reduction in an abundance of the 21cm absorption lines. 
The slope of the likelihood contour becomes flatter for a larger $m_u$ because the smaller ${\delta N}/{N}$ results in a smaller sensitivity on $f_u$.

\begin{table}[h]
\begin{tabular}[t]{c|ccc}
 $\log_{10}(m_u/H_0)$ & 12 & 13 & 13.5  \\
 \hline
 $\Delta f_u$ & 0.05 & 0.07 & 0.13 \\
  \hline
 $\Delta \log_{10}(m_u/H_0)$ & 0.41 & 0.37 & 0.48
\end{tabular}
 \caption{Marginalized errors on $f_u$ and $\log_{10}(m_u/H_0)$ for the
 same fiducial models presented in Fig.~(\ref{fig:ma}). $f_u$ is fixed to $0.2$.}
 \label{table:table1}
\end{table}

\begin{figure*}[h]
 \includegraphics[width=0.3\textwidth]{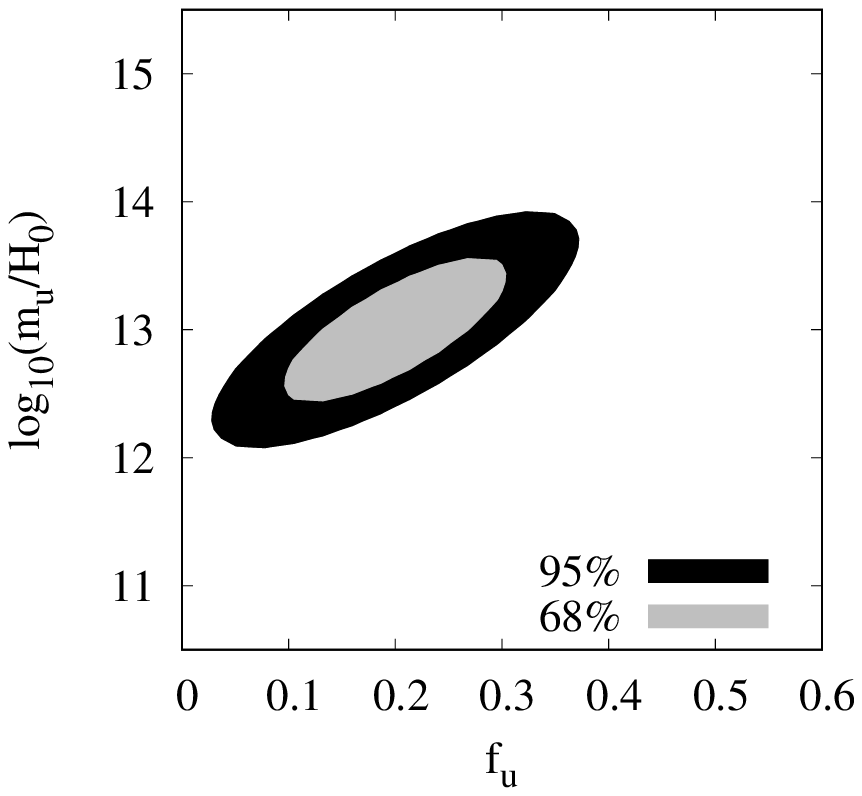}
 \includegraphics[width=0.3\textwidth]{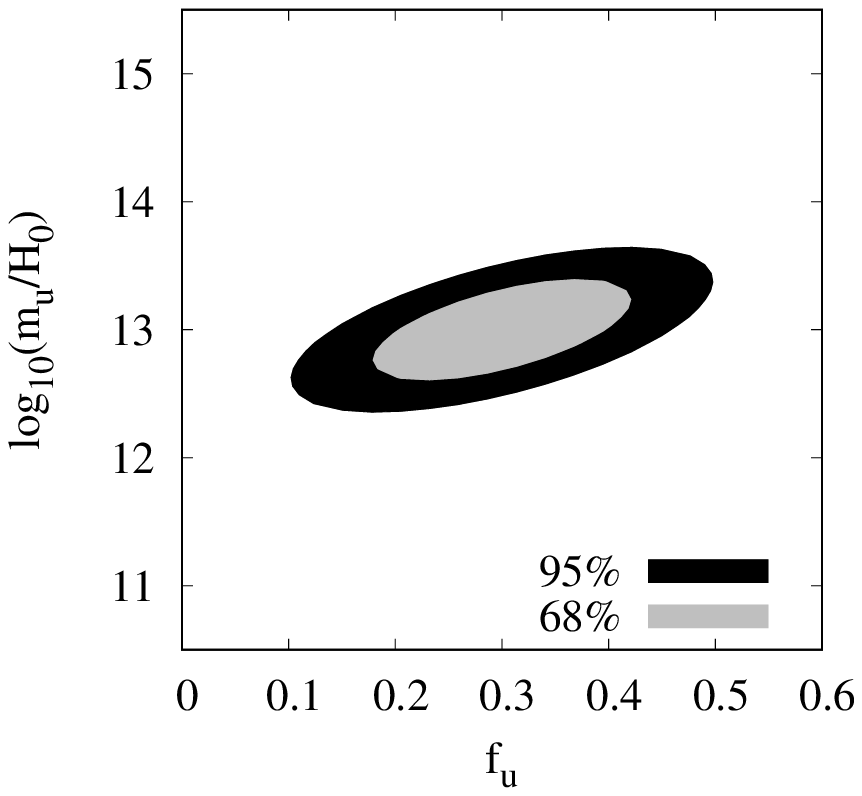}
 \includegraphics[width=0.3\textwidth]{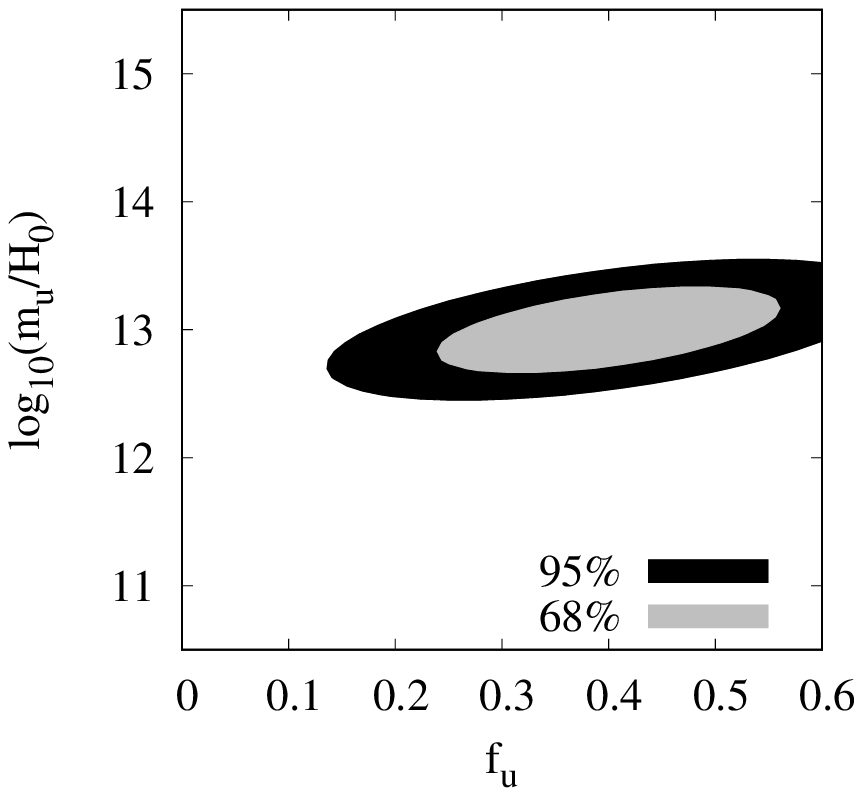}
 \caption{Expected constraints on the $f_u$ and $\log_{10}(m_u/H_0)$
 plane for $f_u=0.2$, $0.3$ and $0.4$ from left to
 right, respectively. The fiducial value of the ULP dark matter fraction
 is fixed to $\log_{10}(m_u/H_0)=13.0$.}  \label{fig:fa}
\end{figure*}

\begin{table}[h]
\begin{tabular}[t]{c|ccc}
 $f_u$ & 0.2 & 0.3 & 0.4  \\
 \hline
 $\Delta f_u$ & 0.07 & 0.08 & 0.11 \\
  \hline
 $\Delta \log_{10}(m_u/H_0)$ & 0.37 & 0.26 & 0.22 
\end{tabular}
 \caption{Marginalized errors on $f_u$ and $\log_{10}(m_u/H_0)$ for the
 same fiducial models presented in Fig.~(\ref{fig:fa}). The mass of ULP is fixed to $\log_{10}(m_u/H_0)=13.0$.}
 \label{table:table2}
\end{table}

In Fig.~\ref{fig:fa}, we show the two dimensional constraints on
$\log_{10}(m_u/H_0)$ and $f_u$ with the fixed fiducial value $\log_{10}(m_u/H_0)=13.0$. We find that the constraints on the mass of ULP becomes tighter as
$f_u$ increases because the ULP with a larger $f_u$ has larger effects on
the matter power spectrum and thus the number of absorption lines. 

In the tables \ref{table:table1} and \ref{table:table2}, we show
the marginalized errors on $f_u$ and $\log_{10}(m_u/H_0)$ for the same
fiducial models presented in Figs.~\ref{fig:ma} and
\ref{fig:fa}. From the tables we find that the pivot scales of the ULP
parameters that the 21cm forest can probe are around $f_u\approx 0.3$
and $\log_{10}(m_u/H_0)\approx 13.0$ where the fractional errors are
minimum.

The reason why the pivot scales arises is as follows.  If $f_u \ll 0.1$,
the amount of suppression in the matter power spectrum due to ULP
becomes negligibly small and it is not possible to place a constraint on
ULP parameters in this limit. If $f_u \gg 0.1$, on the other hand, the
suppression effect becomes too large and we can not have an enough
number of absorption lines for our statistics. As for the mass parameter
$m_u$, the reason is more straightforward. The mass of minihalos that
create 21cm forest absorption lines ranges from $10^{5} M_\odot$ to 
$10^{8} M_\odot$ \cite{2014PhRvD..90h3003S}, which corresponds to the comoving
wavenumber of $30 < k < 1000$ [Mpc$^{-1}$]. 
Therefore the ULP models whose Jeans scale falls within this range can be probed by 21cm forest
observations,  and the Jeans scale for the ULP mass of
$\log_{10}(m_u/H_0)\approx 13$ is in the middle of this range.

Finally, we performed a fisher analysis fixing $f_u=1$ to explore how large a value of $m_u$ can be probed in principle by 21cm forest observations. The results are shown in table \ref{table:table3}. It is shown that 21cm forest observation would have the sensitivity to ULPs with mass as large as $\lesssim 10^{15}H_0$. We find that, however, the effects on the 21cm forest become completely negligible if $\log_{10}(m_u/H_0)>10^{16}$ due to the insufficient abundance in the 21cm absorption lines.

\begin{table}[h]
\begin{tabular}[t]{c|cc}
 $\log_{10}(m_u/H_0)$ & 14 & 15 \\
   \hline
 $\Delta \log_{10}(m_u/H_0)$ & 0.21 & 0.39 
\end{tabular}
 \caption{Errors on $\log_{10}(m_u/H_0)$ fixing $f_u=1$.}
 \label{table:table3}
\end{table}

\begin{figure*}[h]
\begin{minipage}{0.49\linewidth}
 \includegraphics[width=1.0\textwidth]{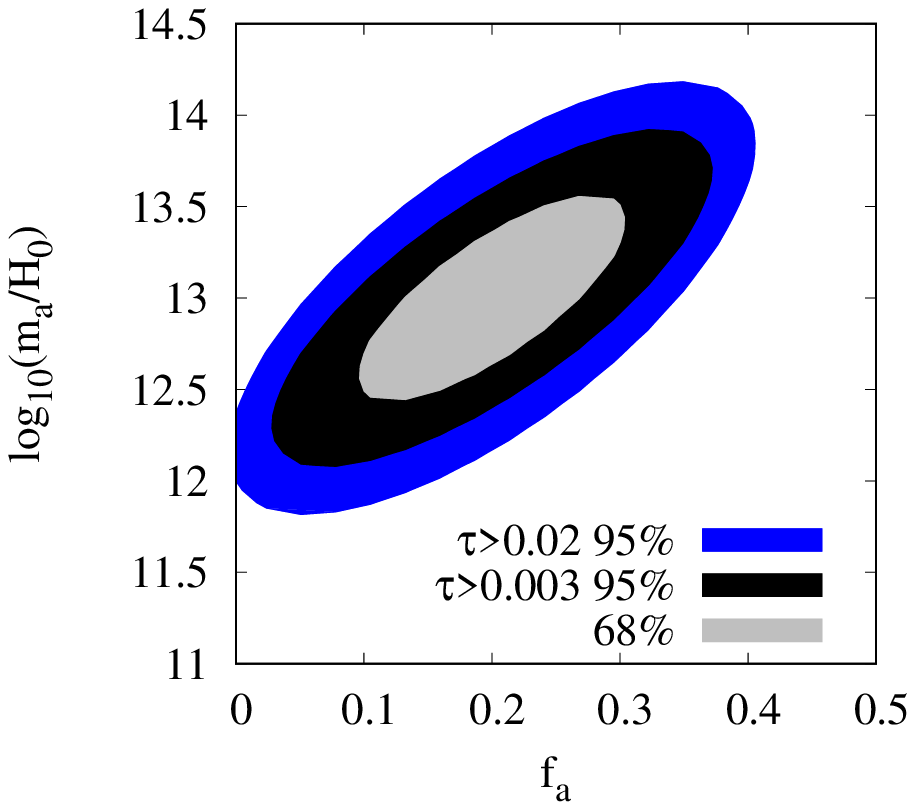}
 \end{minipage}
 \begin{minipage}{0.49\linewidth}
 \includegraphics[width=1.0\textwidth]{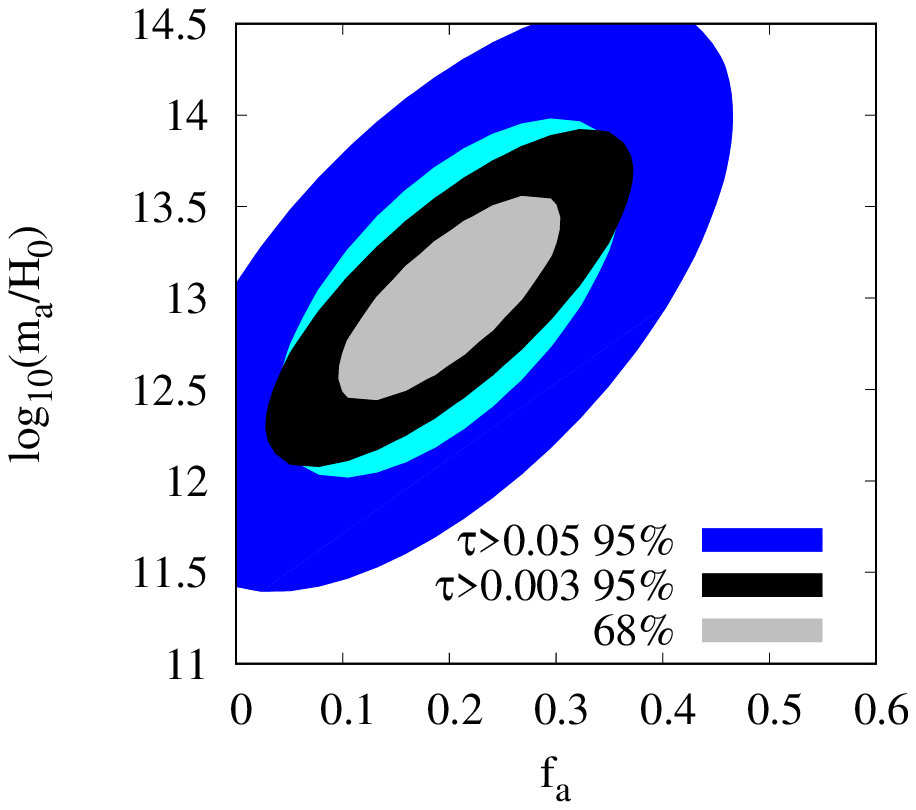}
 \end{minipage}
 \caption{\ichiki{Expected constraints on the $f_a$ and $\log_{10}(m_u/H_0)$ plane for $f_a=0.2$ and $\log_{10}(m_a/H_0)=13.0$, with different thresholds for the optical depth $\tau$ as indicated in the figure.}}  
 \label{fig:pessimistic}
\end{figure*}

\section{Summary \& Discussion}
We studied the bounds on the ultra-light dark matter from the 21cm forest which can potentially probe down to the scale of order kpc, far smaller than the scales accessible by other probes such as the Ly-$\alpha$ probing down to the Mpc scale. Consequently the 21cm potentially can put the tight lower bounds on the ultra-light particles, and we demonstrated that the forthcoming 21cm experiments such as the SKA can probe the mass range up to $m \lesssim 10^{-18}$ eV ($10^{15} H_0$) for $f_u=1$ (more than three orders of magnitude larger than the mass scale probable by the current Ly-$\alpha$ forest observations \cite{2017PhRvL.119c1302I}).
While the effect of the matter power spectrum suppression becomes smaller for $f_u<1$, we also showed that the 21cm forest can probe the ULP mass up to $m=10^{-19}$ eV even if the ULP contribution to the total dark matter density is of order ${\cal O}(10)\%$.

The 21cm forest also has an advantage in using the 21cm absorption spectra from the bright sources and is not susceptible to the foregrounds which give the challenging obstacles in dealing with the 21cm emissions (see for instance Refs. \cite{il2002,seki2014} for the studies on the 21cm emissions from the minihalos), even though the disadvantage is the uncertainty in the existence of the radio loud sources at a high redshift. 

Following \cite{2006MNRAS.370.1867F}, the minimum brightness of the radio background sources required to observe 21cm forest is
\begin{equation}
\begin{split}
S_{\rm min}&=10.4\rm{mJy}\left(\frac{0.01}{\tau}\right)\left(\frac{S/N}{5}\right)\left(\frac{\rm{1 kHz}}{\Delta \nu}\right)^{1/2}\\
 &\times\left(\frac{5000[\rm m^{2}/K]}{A_{\rm{eff}}/T_{\rm{sys}}}\right)\left(\frac{100~\rm{hr}}{t_{{\rm int}}}\right)^{1/2},
\end{split}
\label{eq:Smin}
\end{equation}
where $\tau$ is the target 21cm optical depth, $\nu$ is a frequency resolution, $A_{\mathrm{eff}}$ is an effective collecting area and $T_{\mathrm{sys}}$ is a system temperature, and $t_{\mathrm{int}}$ is the observation time. Assuming the SKA specifications for these quantities, the minimum required flux is of order ${\cal O}(1 \sim 10)$ mJy.
The recent progress and finding of the radio bright sources such as quasars and Gamma-ray bursts (GRBs) at $z\gtrsim 6$ \cite{ban2015,ban2018,2011ApJ...736....7C} warrants the further investigation on the 21cm forest.　For instance, around 10\% of all quasars could be radio loud (radio emission is the dominant component in their spectra) at a high redshift and a radio loud quasar with the flux $8\sim 100$ mJy from 3 GHz to 230 MHz has been found at $z\sim 6$ which can be bright enough for the 21cm forest studies if its local dense gas environment is confirmed by the follow-up surveys \cite{ban2015,ban2018}. \bukuro{The estimates based on extrapolations of the observed radio luminosity functions to the higher redshift indicate that there could be as many as $\sim 10^{4}-10^{5}$ radio quasars with sufficient brightness at $z$=10\cite{Haiman_2004,Xu_2009}.} The GRBs arising from Population (Pop) III stars forming in the metal-free environment are also of great interest because they are expected to be much more energetic objects than ordinary GRBs and thus they can generate much brighter low-frequency radio afterglows, exceeding a tens of mJy\cite{2011ApJ...731..127T}. Recently, the detection of absorption line in the 21cm global signal has been reported \cite{2018Natur.555...67B}, and, if this detection is verified, it implies that Pop III stars should exist at $z\gtrsim 17$\cite{2018MNRAS.480L..43M} as well as Pop III GRBs at $z\gtrsim 10$.

\ichiki{%
In our analyses we have assumed that we are able to count all of the absorption lines with $\tau>0.003$. In reality, however, the level of the thermal noise with a fixed brightness of the radio background sources determines the minimum optical depth
that can be used for the analysis according to Eq.(\ref{eq:Smin}). To see the effect of the minimum optical depth available, we repeated the Fisher analysis setting the minimum optical depth in the integration of Eq.(\ref{eq:tau_integration}) to $\tau_{\rm min} = 0.02$ and $0.05$. We found that, while the size of errors in the case with $\tau_{\rm min}=0.02$ is still comparable to that in the optimistic case with $\tau_{\rm min}=0.003$, in the case of $\tau_{\rm min}=0.05$ the constraint contour inflates as shown in Fig.~(\ref{fig:pessimistic}). Because the distribution of the number of absorption lines has a peak at around $\tau \approx 0.02$ it is of great importance to reduce the noise level to successfully observe the absorption lines with $\tau \gtrsim 0.02$.
}%

\ichiki{%
In addition to the low-level noise, the enough-frequency resolution is necessary for 21cm forest observations. Thermal broadening of the 21cm line at gas temperature $T$ is given by
\begin{equation}
\frac{d\nu}{\nu} = \sqrt{\frac{2kT}{mc^2}} \sim 1.3\times 10^{-5}\left(\frac{T}{10^3 \mbox{[K]}}\right)^{1/2}~,
\label{eq:criteria}
\end{equation}
where we normalized the gas temperature by $10^3$ [K] which is a typical virial temperature of mini-halos. This is the required resolution for 21cm forest observations. The SKA1-low is designed to achieve the frequency resolution as fine as $1$ kHz \cite{2015aska.confE...6C}. The frequency of the 21cm line from redshift $z$ is $1.4/(1+z)$ GHz and therefore SKA1 will achieve $d\nu/\nu = 7.6\times 10^{-6}((1+z)/11)$. Thus, SKA1 will meet the criteria of Eq.(\ref{eq:criteria}) at $z=10$, but the 21cm forest observation may be difficult at higher redshifts.
}%

\begin{figure}
    \centering
    \includegraphics[width=1.0\hsize]{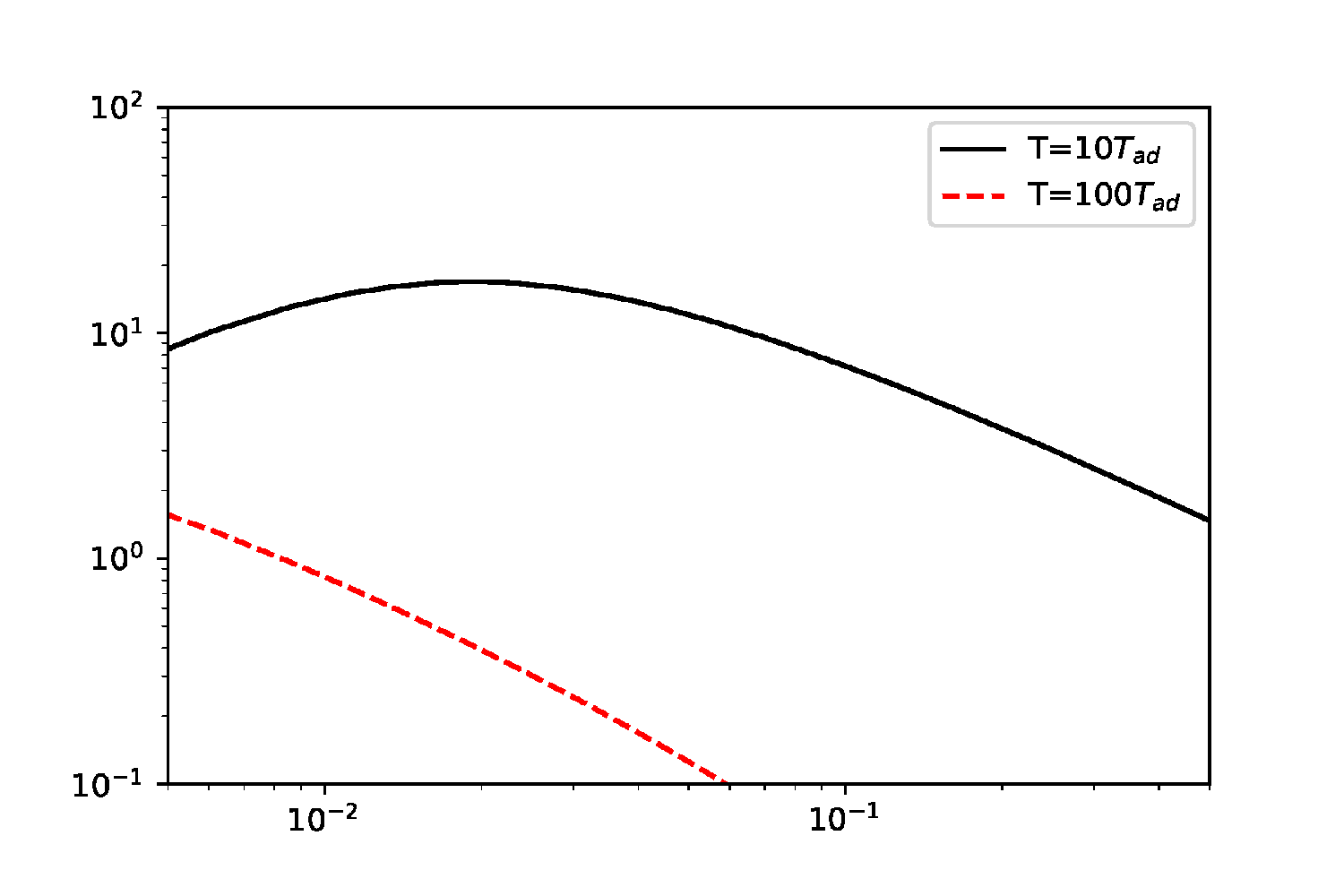}
    \caption{The abundance of 21cm absorption lines for IGM temperature $T=10\times T_{\mathrm{ad}}\sim 20\mathrm{K} ({\it solid})$, $100\times T_{\mathrm{ad}}\sim 200\mathrm{K}$({\it dashed})}
    \label{fig:abundance_T}
\end{figure}

\bukuro{We also comment on the IGM temperature. In this work, we set the IGM temperature as $10 \times T_{\mathrm ad}$, 10 times of the average temperature of the IGM assuming adiabatic cosmic expansion. This value is allowed by recent observation \cite{2015ApJ...809...62P}.  However, X-ray photons as well as shocks driven by supernova explosions, quasar outflows can heat the IGM temperature drastically. Such effects consequently increase the Jeans mass and thus the abundance of the 21cm absorption line is suppressed. Fig.\ref{fig:abundance_T} shows the abundance of the 21cm absorption lines in the case of $T=100T_{\mathrm{ad}}(\sim 200\mathrm{K})$, where the abundance of 21cm absorption lines is suppressed more than an order of magnitude. Furthermore, such astrophysical effects potentially cause the degeneracy between the IGM temperature and ULP properties, and our constraints hence are expected to be weakened in existence of the efficient X-ray heating.}

\kenji{
Even though we implemented the effects of the ultra light dark matter as the suppression of the initial power spectrum analogously to the treatment of the warm dark matter, there have been recent attempts to study how the evolution of structure formation is affected by the quantum pressure \cite{2017PhRvL.119c1302I,moc2017,li2018,moc2019}. Those numerical simulations for instance indicate that the quantum pressure could delay the star formation epoch compared with the warm dark matter scenarios (for instance by the redshift difference of $0.5$ for $m_a\sim 10^{-22}$ eV) hence affecting the estimation of minihalo mass relevant for the 21cm forest observations. While the numerical simulations properly treating the detailed properties of the ultra-light dark matter are computationally demanding and still under the scrutiny, we leave the further study including the dynamical effects of the quantum pressure on the structure formation and their indications for the 21cm forest signals for the future work.
}

The 21cm forest appearing in radio background spectra at a high redshift would give a promising probe on the nature of the dark matter and the epoch of reionization which remain among the most crucial open questions in cosmology.

\begin{acknowledgments}
HS is supported by the NSFC (Grant No.11850410429), the China Postdoctoral Science Foundation, the Tsinghua International Postdoctoral Fellowship Support Program, and the International Postdoctoral Fellowship from the Ministry of Education and the State Administration of Foreign Experts Affairs of China. KK is supported by the Institute for Basic Science (IBS-R018-D1). KI was supported in part by JSPS KAKENHI Grant Numbers 18K03616 and 17H01110.
\end{acknowledgments}

\nocite{*}

\bibliography{apssamp}

\end{document}